\def\prd{Phys. Rev. D}

\def\apj{Astrophys. J.}

\def\mnras{Mon. Not. R. Astr. Soc.}
\def\aap{Astr. Astrophys.}

\def\jcap{JCAP}

\def \<{\langle}
\def \>{\rangle}

\newcommand{\ra}{\;\raise1.0pt\hbox{$'$}\hskip-6pt\partial\;}
\newcommand{\lo}{\;\overline{\raise1.0pt\hbox{$'$}\hskip-6pt\partial}\;}

\newcommand{\degree}{^\circ}

\newcommand{\Abs}{\abstract}
\newcommand{\Ack}{\acknowledgments}
\newcommand{\mktt}{\maketitle}

\documentclass[a4paper,11pt]{article}
\usepackage{jcappub,graphicx,epsfig,natbib,color,times,bm,amsmath,multirow,hyperref}
\usepackage[ddmmyyyy,hhmmss]{datetime}

\begin{document}

\title{Blind correction of the EB-leakage in the pixel domain}

\author[a,b]{Hao Liu,}\emailAdd{liuhao@nbi.dk}

\author[a]{James Creswell,}\emailAdd{james.creswell@nbi.ku.dk}

\author[a]{Konstantina Dachlythra}\emailAdd{kwn.dachlythra@gmail.com}

\affiliation[a]{The Niels Bohr Institute \& Discovery Center, Blegdamsvej 17,
DK-2100 Copenhagen, Denmark}

\affiliation[b]{Key Laboratory of Particle and Astrophysics, Institute of High
Energy Physics, CAS, 19B YuQuan Road, Beijing, China}

\Abs{

We study the problem of EB-leakage that is associated with incomplete
polarized CMB sky. In the blind case that assumes no additional information
about the statistical properties and amplitudes of the signal from the missing
sky region, we prove that the recycling method (Liu et al.~2018) gives the
unique best estimate of the EB-leakage. Compared to the previous method, this
method reduces the uncertainties in the BB power spectrum due to EB-leakage by
more than one order of magnitude in the most interesting domain of multipoles,
where $\ell$ is between $80$ and $200$. This work also provides a useful
guideline for observational design of future CMB experiments.

}


\mktt

\section{Introduction}\label{sec:intro}

The most promising way to detect primordial gravitational waves is by
measuring the B-mode polarization of the Cosmic Microwave Background (CMB)
radiation. This requires first separating the B-mode polarization from the
dominant E-mode polarization. However, because the E/B decomposition is
non-local, when only part of the sky is visible, the calculated B-mode is
corrupted by power originating in the E-mode, which is called the E-to-B
leakage~\cite{2002PhRvD..65b3505L, Bunn:2002df}. Since neither current
experiments nor those upcoming in the next few decades will be able to provide
reliable full sky background data, the E-to-B leakage is a problem that must
be solved before one can detect primordial gravitational waves. In this paper
we show how to make the best blind estimate (BBE) of this E-to-B leakage using
the data within the available sky region and the shape of this region.

Depending on the prior assumptions and the starting point, there are three
types of estimations: the blind, prior, and posterior estimations. Blind
estimation means there are no prior assumptions, while the other two are based
on the context of data $\bm{x}$ described by a model with parameters
$\bm{\Theta}$. Prior estimation means to make an estimate of the data from
model parameters, and posterior estimation means to estimate the model
parameters from the data. In practice, these two concepts are connected
through Bayes' theorem $P(\bm{x}|\bm{\Theta}) P(\bm{\Theta}) =
P(\bm{\Theta}|\bm{x})P(\bm{x})$, where $P(\bm{x}|\bm{\Theta})$ is the
conditional probability of $\bm{x}$ with given $\bm{\Theta}$, and
$P(\bm{\Theta} | \bm{x})$ is the conditional probability of $\bm{\Theta}$ with
given $\bm{x}$. Although posterior estimation is only found in the context of
parameter estimation because it targets model parameters, the idea of prior
estimation can be extended to include a more general notion of estimation with
prior constraints, and does not necessarily require a physical model.

Strictly speaking, posterior estimation is unsuitable for the EB-leakage
problem, because the goal is the data $\bm{x}$, not model parameters
$\bm{\Theta}$. Indirect usage of a posterior estimation might be possible, but
several problems have to be solved in advance, which will be discussed in
section~\ref{sec:fisher}. Besides the posterior estimation, we have two other
options for the EB-leakage problem: either blind estimation or prior
estimation. This paper focuses on the blind estimation, and the prior
estimation will be studied in a future work.

Normally, the best estimation is defined to be the unbiased estimation with
the smallest error. However, in the case of blind estimation of EB-leakage,
where we allow no constraints on the unavailable sky region, the error of the
estimation is completely undetermined and cannot be presented in the form of
error bars. Therefore, the BBE is defined as follows:

Let $\bm{S}(\bm{p}_1,\cdots,\bm{p}_n,\bm{q}_1,\cdots,\bm{q}_m)$ be the real
EB-leakage, where $\bm{p}_i$ are the available pixels, and $\bm{q}_i$ are the
unavailable pixels. If $\bm{S}$ can be decomposed into
\begin{equation}\label{equ:exp}
\bm{S}(\bm{p}_1,\cdots,\bm{p}_n,\bm{q}_1,\cdots,\bm{q}_m) = 
\bm{E}(\bm{p}_1,\cdots,\bm{p}_n)+\bm{\Delta}(\bm{q}_1,\cdots,\bm{q}_m) + 
\mathrm{const},
\end{equation}
then $\bm{E}(\bm{p}_1, \cdots,\bm{p}_n)$ is the BBE, and
$\bm{\Delta}(\bm{q}_1, \cdots,\bm{q}_m)$ is the error. 

The definition of the BBE means that the error is a function only of the
missing sky pixels, and the BBE depends on the available sky pixels in exactly
the same way as the true leakage does. Therefore, \textbf{any further
improvement of the BBE requires additional information about the missing sky
region}. With the Taylor series expansion of $\bm{S}$, one can easily prove
that, if the decomposition in eq.~(\ref{equ:exp}) exists, then
$\bm{E}(\bm{p}_1,\cdots,\bm{p}_n)$ is unique, thus the BBE is unique except
for a constant offset.\footnote{Note that the true EB-leakage $\bm{S}$ is not
unique with missing data, as described by $\bm{\Delta}$. Only $\bm{E}$ is
unique.} In this work, we not only give the mathematical form of $\bm{E}$, but
also point out how to calculate it efficiently.

In ref.~\cite{2018arXiv181104691L}, we introduced the background of the
EB-leakage problem, and used the ``recycling method'' to correct the E-to-B
leakage. The method is fast, simple, and provides much better results than
previous corrections. In this work, we prove that the recycling method gives
exactly the BBE of EB-leakage. Meanwhile, it should also be noted that the
detection of the primordial B-mode is very complicated -- as mentioned in
ref.~\cite{2018arXiv181104691L}, there are at least five main obstacles:
foreground removal, delensing, noise, systematics, and the $EB$ leakage. A
successful solution of the EB-leakage is just one step of the whole effort.

This paper is organized as follows: in section~\ref{sec:notation}, we
introduce the basis and notations, and in section~\ref{sec:proof} we prove
that the recycling method gives the BBE of EB-leakage, and explicitly give the
analytic form of the BBE. The possibility of further correction with
additional information is discussed in section~\ref{sec:fisher}, and the
conclusion is given in section~\ref{sec:conclusion}.

\section{Basis and notations}\label{sec:notation}

We briefly review the calculation of $E$- and $B$-family maps by pixel domain
convolution. More details can be found in refs.~\cite{2018JCAP...05..059L,
Rotti:2018pzi, 2018A&A...617A..90L}.

Given a polarized sky map $\bm{P}(\bm{n})= (Q(\bm{n}),U(\bm{n}))$, the true $E$-
and $B$-family maps are:
\begin{align}\begin{split}\label{equ:EB-QU_final simple}
\bm{P}_E(\bm{n}) = 
  \begin{pmatrix} Q_{E} \\ U_{E} \end{pmatrix}(\bm{n}) 
  =& \int G_E(\bm{ n},\bm{ n}') \bm{P}(\bm{n}')\, d \bm { n}', \\
\bm{P}_B(\bm{n}) = 
  \begin{pmatrix} Q_{B} \\ U_{B} \end{pmatrix}(\bm{n}) 
  =& \int G_B(\bm{ n},\bm{ n}') \bm{P}(\bm{n}')\, d \bm { n}', \\
\end{split}\end{align}
with $\bm{P}_E(\bm{n})+ \bm{P}_B(\bm{n}) = \bm{P}(\bm{n}) $ and:
\begin{align}\begin{split}\label{equ:EB-QU_final def2}
G_E(\bm{ n},\bm{ n}') = &
\begin{pmatrix}
G_1  & +G_2 \\
+G_3  & G_4
\end{pmatrix}
(\bm{ n},\bm{ n}'), \\
G_B(\bm{ n},\bm{ n}') = &
\begin{pmatrix}
G_4  & -G_3 \\
-G_2  & G_1
\end{pmatrix}
(\bm{ n},\bm{ n}').
\end{split}\end{align}
The $G_{1-4}$ functions are defined as:
\begin{align}\begin{split}
\label{equ:define G}
G_{1}(\bm{ n},\bm{ n}') &= 
  \sum_{\ell,m} F_{+,\ell m}(\bm{n})F^*_{+,\ell m}(\bm{ n}'), \\\quad 
G_{2}(\bm{ n},\bm{ n}') &= 
  \sum_{\ell,m} F_{+,\ell m}(\bm{n})F^*_{-,\ell m}(\bm{ n}'), \\
G_{3}(\bm{ n},\bm{ n}') &= 
  \sum_{\ell,m} F_{-,\ell m}(\bm{n})F^*_{+,\ell m}(\bm{ n}'), \\\quad 
G_{4}(\bm{ n},\bm{ n}') &=
  \sum_{\ell,m} F_{-,\ell m}(\bm{ n})F^*_{-,\ell m}(\bm{ n}'),
\end{split}\end{align}
and the $F_{+,-}$ functions are defined in terms of the spin-2 spherical
harmonics as:
\begin{align}\begin{split}
\label{equ:define F}
F_{+,\ell m}(\bm{ n}) &= -\frac{1}{2} \left[{}_{2}Y_{\ell m}(\bm{ n}) +
  {}_{-2}Y_{\ell m}(\bm{ n})\right], \\
F_{-,\ell m}(\bm{ n}) &= -\frac{1}{2i}\left[{}_{2}Y_{\ell m}(\bm{ n}) - 
  {}_{-2}Y_{\ell m}(\bm{ n}) \right].
\end{split}\end{align}
Note that $G_i$ are real, $G_2 = G_3$,  and $G_1 + G_4 = \delta$. 

As an augmentation, the $G_E$ and $G_B$ kernels can be written in terms of a
common kernel $G$ and a delta function:
\begin{align} \begin{split}
\label{eq:Gdelta}
G_E(\bm{n}, \bm{n}') &= \frac{1}{2} \delta(\bm{n} - \bm{n'}) + 
  G(\bm{n}, \bm{n}'),\\
G_B(\bm{n}, \bm{n}') &= \frac{1}{2} \delta(\bm{n} - \bm{n'}) - 
  G(\bm{n}, \bm{n}').
  \end{split}
\end{align}
In practice, when applied to pixelized sky maps, the sums in
eq.~(\ref{equ:define G}) are not taken to $\ell = \infty$ but instead to a
finite $\ell_\mathrm{max}$. In this case, the identities $G_2 = G_3$ and $G_1
+ G_4 = \delta$ are broken and the delta functions in eq.~(\ref{eq:Gdelta})
are replaced by ``bandpassed'' delta functions, which behave similarly but
maintain the orthogonality of the $E$ and $B$ modes for any finite
$\ell_\mathrm{max}$.

For convenience, the operation of extracting $\bm{P}_E(\bm{n})$ or
$\bm{P}_B(\bm{n})$ from $\bm{P}(\bm{n})$ using eq.~(\ref{equ:EB-QU_final
simple}) is written in form of operators as follows:
\begin{align}\begin{split}\label{equ:operator}
\Psi_E(\bm{P}) &\Rightarrow \bm{P}_E(\bm{n}), \\ 
\Psi_B(\bm{P}) &\Rightarrow \bm{P}_B(\bm{n}).
\end{split}
\end{align}

\section{Proof of best correction}\label{sec:proof}

In this section, we give the mathematical form of the BBE of EB-leakage, and
point out how to calculate it efficiently in practice.

Given a sky mask $\bm{M}(\bm{n})$, which takes values of 1 or 0 depending on
whether $\bm{n}$ is available or not, and use the notation in
eq.~(\ref{equ:operator}), the true EB-leakage is
\begin{align}\begin{split}\label{equ:real leakage}
\bm{L}(\bm{n})_{\mathrm{true}} &= \Psi_B(\bm{M}\Psi_E(\bm{P})) \\
&= \int G_B(\bm{ n},\bm{ n}')\bm{M}(\bm{n'})\bm{P_E}(\bm{ n}')\,d \bm {n}'\\
&= \int G_B(\bm{ n},\bm{ n}')\bm{M}(\bm{n'}) \left[\int G_E(\bm{n}',\bm{n}'') 
  \bm{P}(\bm{n}'')\, d \bm {n}''\right]\,d \bm {n}'\\
&= \int \bm{P}(\bm{n}'')\, d \bm {n}''\int G_B(\bm{ n},\bm{ n}') 
  G_E(\bm{n}',\bm{n}'')\bm{M}(\bm{n'})\,d \bm {n}' \\
&= \int G_{EB}(\bm{ n},\bm{ n}'') \bm{P}(\bm{n}'')\, d \bm {n}'',
\end{split}\end{align}
where
\begin{align}\begin{split}\label{equ:EB-kernel}
G_{EB}(\bm{ n},\bm{ n}'') \equiv \int G_B(\bm{ n},\bm{ n}') 
  G_E(\bm{n}',\bm{n}'')\bm{M}(\bm{n'})\,d \bm {n}' 
\end{split}\end{align}
is the EB-leakage convolution kernel. It is fully determined by $\bm{M}(\bm{n'})$.
If $\bm{M}(\bm{n'})=1$ (no mask), then $G_{EB}(\bm{ n},\bm{ n}'') = 0$.

Given a mask, the EB-leakage is affected by outside-to-inside propagation and
vice versa, which makes the estimation complicated. Eq.~(\ref{equ:real
leakage}) gives the leakage as a convolution of the kernel $G_{EB}(\bm{
n},\bm{ n}'')$ and the input sky map $\bm{P}(\bm{n}'')$. The kernel describes
the full propagation effect without involving the particular sky map. This
clear separation makes the following study much easier: when a mask is present
and given no additional information of the missing sky region, all available
information is fully described by $\bm{M}(\bm{n''})\bm{P}(\bm{n}'')$, hence
the BBE of EB-leakage is
\begin{eqnarray}\label{equ:leakage max}
\bm{L}(\bm{n})_{\mathrm{best}}= \int G_{EB}(\bm{ n},\bm{ n}'')\bm{M}(\bm{n''})
  \bm{P}(\bm{n}'')\, d \bm {n}''.
\end{eqnarray}
Eq.~(\ref{equ:leakage max}) can also be understood through the error of
$\bm{L}(\bm{n})_{\mathrm{best}}$:
\begin{eqnarray}\label{equ:leakage diff}
\bm{L}(\bm{n})_\mathrm{error}
= \bm{L}(\bm{n})_\mathrm{true}-\bm{L}(\bm{n})_\mathrm{best} 
= \int G_{EB}(\bm{ n},\bm{ n}'')[1-\bm{M}(\bm{n''})]\bm{P}(\bm{n}'')\, d \bm {n}'',
\end{eqnarray}
which is 100\% determined by the missing sky region. Therefore,
\textbf{without additional information about the missing sky region, it is
impossible to reduce} $\bm{L}(\bm{n})_\mathrm{error}$. This fully satisfies
the definition of the BBE in section~\ref{sec:intro}.

Eqs.~(\ref{equ:real leakage}) and~(\ref{equ:leakage max}) are difficult to
calculate directly. However, using the recycling method from
ref.~\cite{2018arXiv181104691L}, it is possible to calculate
$\bm{L}(\bm{n})_\mathrm{template}$ instead of $\bm{L}(\bm{n})_\mathrm{best}$,
which does not need the computationally expensive kernel $G_{EB}(\bm{ n},\bm{
n}'')$:
\begin{align}
\begin{split}\label{equ:leakage template}
&\bm{L}(\bm{n})_\mathrm{template} = 
  \int G_B(\bm{ n},\bm{ n}')\bm{M}(\bm{n'})\bm{P}_E'(\bm{ n}')\,d \bm {n}'\\
&= \int G_B(\bm{ n},\bm{ n}')\bm{M}(\bm{n'})\left[\int G_E(\bm{n}',\bm{n}'') 
  \bm{M}(\bm{n''})\bm{P}(\bm{n}'')\, d \bm {n}''\right]\,d \bm {n}'\\
&= \int \bm{M}(\bm{n''})\bm{P}(\bm{n}'')\, d \bm {n}''\int G_B(\bm{ n},\bm{ n}') 
  G_E(\bm{n}',\bm{n}'')\bm{M}(\bm{n'})\,d \bm {n}' \\
&= \int G_{EB}(\bm{ n},\bm{ n}'')\bm{M}(\bm{n''}) \bm{P}(\bm{n}'')\, d \bm {n}'' \\
&= \bm{L}(\bm{n})_\mathrm{best}.
\end{split}
\end{align}

Because $\bm{L}(\bm{n})_\mathrm{template}$ and $\bm{L}(\bm{n})_\mathrm{best}$
are identical, the template from the recycling method is exactly the BBE of
EB-leakage. As shown by eq.~(\ref{equ:leakage template}),
$\bm{L}(\bm{n})_\mathrm{best}$ can be easily calculated by two steps: 1) Apply
the mask to $\bm{P}(\bm{n})$ and get $\bm{P}_E'$. 2) Apply the mask to
$\bm{P}_E'$ and get $\bm{P}_B''$ as the leakage estimate. Using the notations
in eq.~(\ref{equ:operator}), the final form of the BBE is the following:
\begin{eqnarray}
\bm{L}(\bm{n})_\mathrm{best} = \Psi_B(\bm{M}\Psi_E(\bm{MP})).
\end{eqnarray}

\section{Possibility of further improvement with additional information}
\label{sec:fisher}

All the above analysis is blind and makes no assumptions about the missing sky
signal, not even Gaussianity or isotropy. In this case, the recycling method
gives the BBE of EB-leakage.

Given additional information, it might be possible to partly reconstruct the
missing sky region, e.g., using lossless Fisher
estimators~\cite{1997PhRvD..55.5895T}, as was done in
ref.~\cite{2010MNRAS.407.2530E} for the temperature case. If this can be done
properly for polarized maps, then the EB-leakage estimation can certainly be
improved. However, there is an important constraint that was repeatedly
mentioned in refs.~\cite{1997PhRvD..55.5895T, 2010MNRAS.407.2530E} and other
works, that both Gaussianity and isotropy have to be assumed for current
Fisher estimators, because only then can the statistical properties of the
covariance matrix be fully determined by the angular power spectrum.

Unfortunately, the EB-leakage is highly non-isotropic, and therefore cannot be
estimated using current Fisher estimators. It could be possible to redesign
the Fisher estimator and remove the requirement for isotropy. However, several
difficulties must be solved: for example, the covariance matrix is
non-analytic and is always singular (due to the missing region), and the map
of a single component (like the EB-leakage alone) is not available at the
beginning of estimation. Alternatively, instead of the posterior Fisher
estimator, one could use a prior estimator that incorporates given prior
information. This approach will be investigated in future work. However, some
simple assumptions can be made that give minor, though immediate,
improvements, such as assuming that $\bm{L}(\bm{n})_\mathrm{best}$ is
uncorrelated with $\bm{P}_B(\bm{n})$, in which case the correction can be
slightly improved by removing the template using linear fitting. This was
adopted in ref.~\cite{2018arXiv181104691L}.

\section{Conclusion and discussion}\label{sec:conclusion}

\begin{figure*}[!htb]
    \centering
    \includegraphics[width=0.8\textwidth]{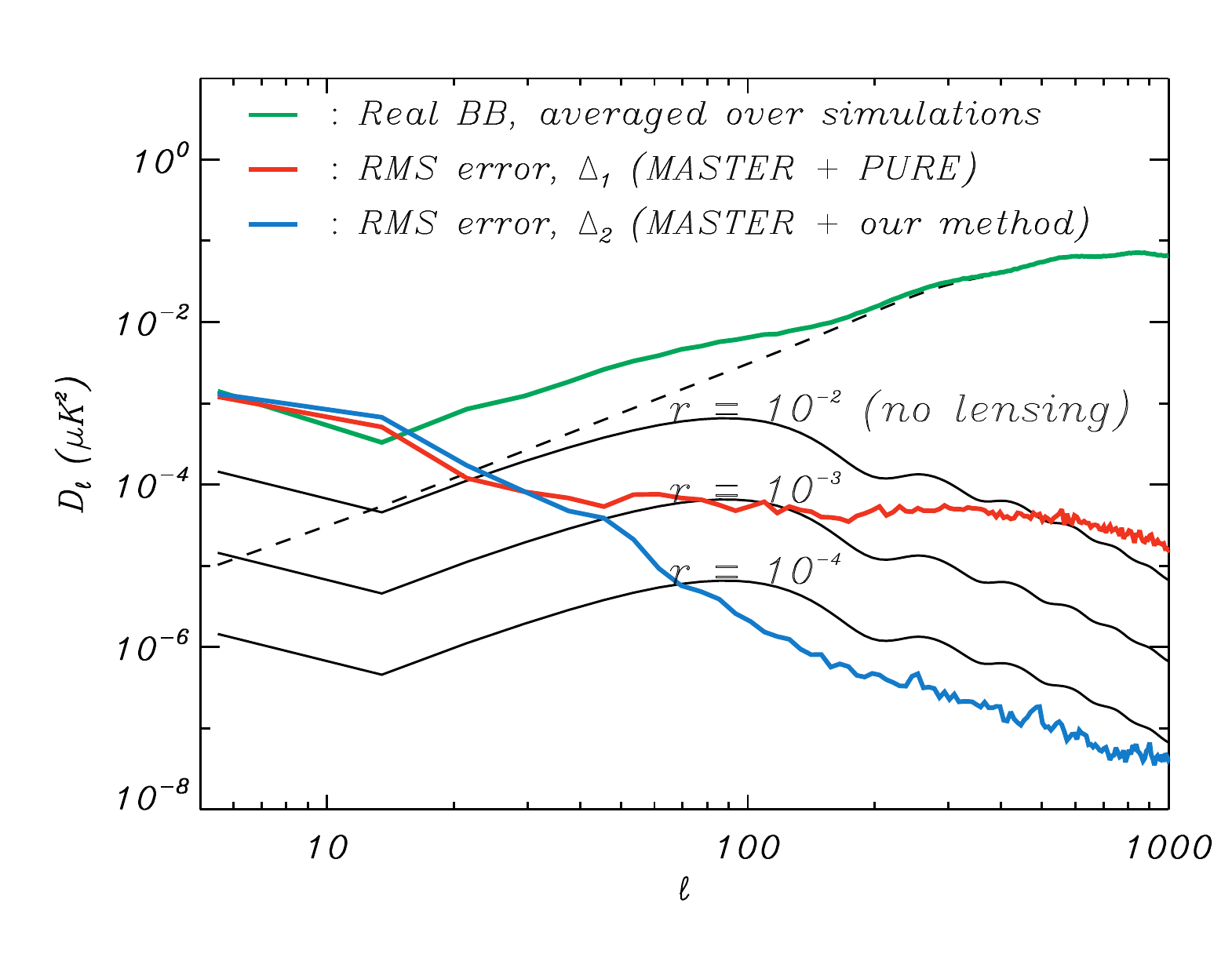}
    \caption{Comparison of the errors of $EB$~leakage correction: red for
    MASTER+PURE and blue for MASTER+our method. The primordial $B$-mode
    spectra for $r=10^{-2}\sim10^{-4}$ (black solid) and the lensing $B$-mode
    spectrum (black dashed) are also added for comparison. Everything is done
    under the same conditions (resolution, simulated maps, sky region,
    apodization, etc.). This is similar to figure 5 of
    ref.~\cite{2018arXiv181104691L}, with only one difference that the linear
    fitting procedure is skipped to make the estimation completely blind.}
    \label{fig:with master}
\end{figure*}

In this work, by proving that the recycling method gives the BBE of
EB-leakage, the problem of EB-leakage is completely solved in the blind case.
To illustrate the correction method, we run a test that is similar to the one
in figure~5 of ref.~\cite{2018arXiv181104691L}, differing only in that we skip
the linear fitting procedure (see section~\ref{sec:fisher}) to make the
estimation completely blind. In the test, we calculate the EB-leakage
correction using either our recycling method or the
PURE-method~\cite{Bunn:2002df, 2003PhRvD..68h3509L, 2003NewAR..47..987B,
2004mmu..symp..309Z, PhysRevD.82.023001, 2006PhRvD..74h3002S,
2007PhRvD..76d3001S}. The PURE method is also blind and was previously the
best one. In both cases, the full sky B-mode spectrum is reconstructed using
the MASTER method~\citep{2002ApJ...567....2H} and the \textsc{NaMaster}
code~\citep{Alonso:2018jzx, namaster}. The results are shown in
figure~\ref{fig:with master}, which is almost the same as figure~5 of
ref.~\cite{2018arXiv181104691L}: the correction is 1--2 orders of magnitude
better than the PURE method, and in the most important multiple range for
detecting primordial gravitational waves, i.e. $80\le\ell\le200$, our result
is good enough to detect $r\approx10^{-4}$, which is sufficient for the next
few decades. Comparison with figure~5 of ref.~\cite{2018arXiv181104691L} shows
that the linear fitting used in ref.~\cite{2018arXiv181104691L} helps to
improve the result by about $40\%$. This nicely illustrates the final option
mentioned in section~\ref{sec:fisher}: the improvement is not big, but still
good because it costs almost nothing.

As mentioned in ref.~\cite{2018arXiv181104691L}, the five main obstacles in
detection of CMB $B$-modes are foreground removal, delensing, noise,
systematics, and the EB-leakage. In the blind case, this work reduces the list
to four, which is a solid step towards real detection of the primordial
gravitational waves.

This work is also relevant for the design of future CMB experiments: if there
is no additional information about the missing region, then the ability to
detect $r$ in a certain observation region is limited by the BBE of EB-leakage
for that region. Therefore it is unnecessary to increase the detector
sensitivity to an extent that significantly exceeds this limit. In order to be
sensitive to even lower $r$, it would be necessary to enlarge the region of
observation, which will probably also require longer observing time and more
accurate foreground removal.

\Ack{

We sincerely thank Sebastian von Hausegger and Pavel Naselsky for very helpful
discussions. This research has made use of the
\textsc{HEALPix}~\citep{2005ApJ...622..759G} package and the
\textsc{NaMaster}/\textsc{pymaster} package~\citep{namaster,Alonso:2018jzx},
and was partially funded by the Danish National Research Foundation (DNRF)
through establishment of the Discovery Center and the Villum Fonden through
the Deep Space project. Hao Liu is also supported by the National Natural
Science Foundation of China (Grants No. 11653002, 1165100008), the Strategic
Priority Research Program of the CAS (Grant No. XDB23020000) and the Youth
Innovation Promotion Association, CAS. \\

}

\appendix

\section{Examples of the EB convolution kernel}\label{sec:kernel}

The convolution kernel $G_{EB}(\bm{ n},\bm{ n}'')$ can be calculated from
eq.~(\ref{equ:EB-kernel}). Compared to direct calculation (which is quite
difficult), a more convenient way is to do it from eq.~(\ref{equ:real
leakage}) using a Dirac delta function:
\begin{align}\begin{split}\label{equ:kernel compute}
&G_{EB}(\bm{ n},\bm{n}_0) = \bm{L}(\bm{n})_{\mathrm{true},\bm{n}_0} = \\
&= \int G_B(\bm{ n},\bm{ n}')\bm{M}(\bm{n'}) \left[\int G_E(\bm{n}',\bm{n}'') 
  \bm{P}(\bm{n}'')\delta(\bm{n}''-\bm{n}_0)\, d \bm {n}''\right]\,d \bm {n}'\\
&= \int G_{EB}(\bm{ n},\bm{ n}'') \bm{P}(\bm{n}'')
  \delta(\bm{n}''-\bm{n}_0)\, d \bm {n}'',
\end{split}\end{align}

In practice, eq.~(\ref{equ:kernel compute}) means to obtain $G_{EB}(\bm{
n},\bm{n}_0)$ as follows:
\begin{enumerate}
\item \label{itm:kernel step1}Start from a zero map and set $Q(\bm{n}_0)$ or
$U(\bm{n}_0)$ to 1.
\item \label{itm:kernel step2}Calculate $\bm{P}_E$ without mask.
\item \label{itm:kernel step3}Calculate $\bm{P}_B$ from the output of
step~\ref{itm:kernel step2} with a mask.
\end{enumerate}

Steps~\ref{itm:kernel step1}--\ref{itm:kernel step3} give $G_{EB}(\bm{
n},\bm{n}_0)$. However, $G_{EB}(\bm{n}_0,\bm{ n})$ is easier to understand,
because the EB-leakage at $\bm{n}_0$ (point of interest) is simply
\begin{equation}
\bm{L}(\bm{n}_0)_\mathrm{true}=\int G_{EB}(\bm{n}_0,\bm{n})\bm{P}(\bm{n})\,d\bm{n}.
\end{equation}
This can be done by repeating steps~\ref{itm:kernel step1}--\ref{itm:kernel
step3} for all possible $\bm{n}_0$, and deriving $G_{EB}(\bm{n}_0,\bm{ n})$
from the results. The calculation is time consuming, and here we present the
results only for low resolution.

\begin{figure}
  \centering
  \includegraphics[width=0.32\textwidth]{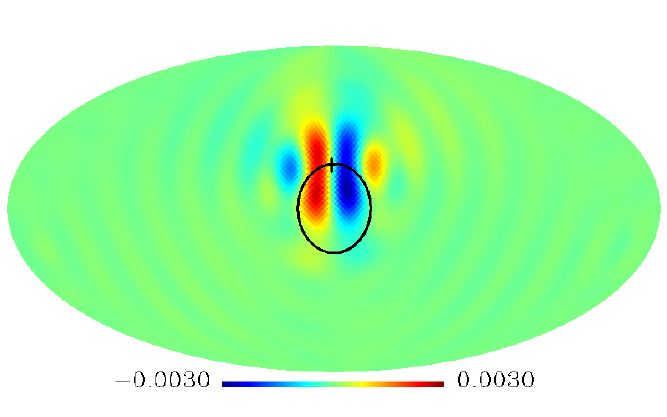}
  \includegraphics[width=0.32\textwidth]{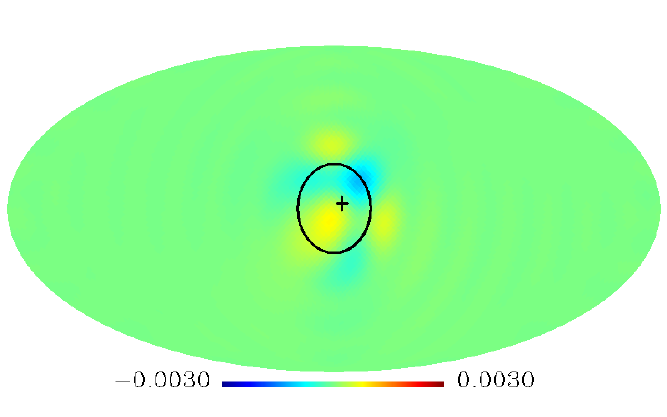}
  \includegraphics[width=0.32\textwidth]{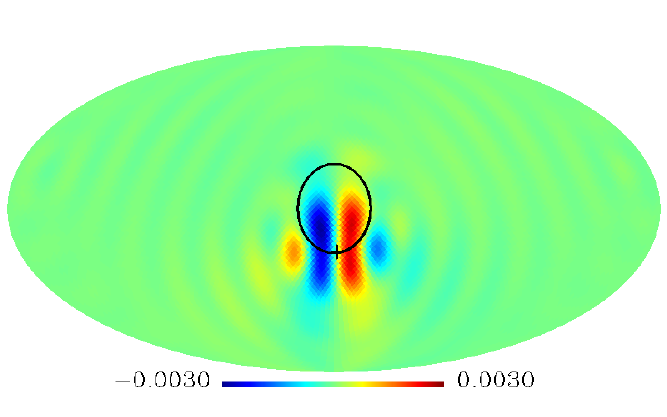}
  
  \includegraphics[width=0.32\textwidth]{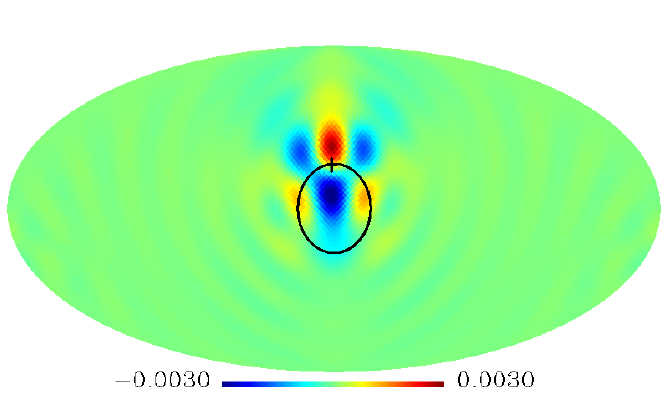}
  \includegraphics[width=0.32\textwidth]{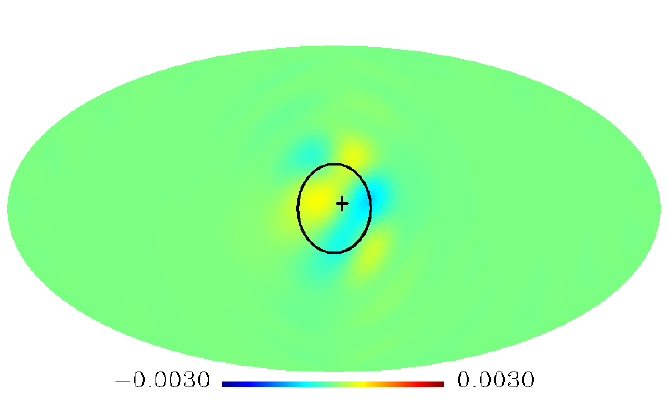}
  \includegraphics[width=0.32\textwidth]{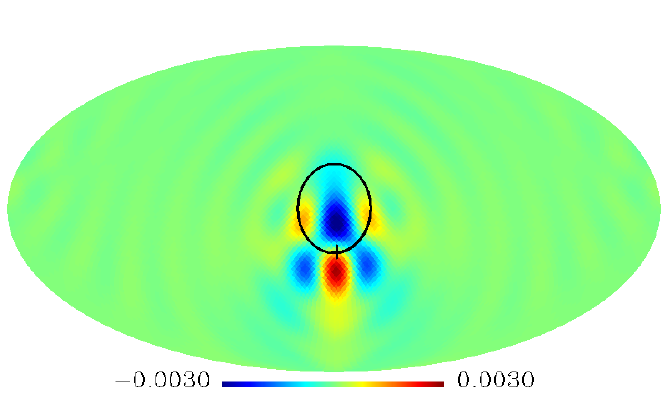}
  \caption{Examples of the EB-convolution kernel $G_{EB}(\bm{n}_0,\bm{ n})$.
  Cross for $\bm{n}_0$ (fixed in each example) and circle for the edge of the
  mask. \emph{Upper}: $U=0$. \emph{Lower}: $Q=0$. $N_{side}=32$ and
  $\ell_{max}=16$.}
  \label{fig:eb kernel}
\end{figure}

In figure~\ref{fig:eb kernel}, we show examples of $G_{EB}(\bm{n}_0,\bm{ n})$
for $N_{side}=32$ and $\ell_\mathrm{max}=16$. The mask is a $r=20\degree$ disk mask
located in the center of the map. One can see that when $\bm{n}_0$ (cross) is
in the middle of the mask, the kernel is relatively much weaker than when
$\bm{n}_0$ is at the edge, which is consistent to the well-known fact that the
center region contains much less EB-leakage than the edge region.

\providecommand{\href}[2]{#2}\begingroup\raggedright\endgroup

\end{document}